\documentclass[twocolumn, secnumarabic,amssymb, nobibnotes, aps, prl, superscriptaddress]{revtex4-2}

\usepackage[colorlinks, linkcolor=blue, anchorcolor=blue, allcolors=blue, citecolor=blue]{hyperref}
\usepackage{amssymb}
\usepackage{amsthm}
\usepackage{graphicx}
\usepackage{amsmath}
\usepackage{array}
\usepackage{booktabs}
\usepackage{color}
\usepackage[normalem]{ulem}
\usepackage{bm}

\begin{document}

\title{Resetting a Superconducting Qubit with a Compact Colder Resistive Dissipator below the Qubit Frequency}

\author{Jiajia Tian}
\thanks{These authors contributed equally to this work.}
\affiliation{Beijing Key Laboratory of Fault-Tolerant Quantum Computing, Beijing Academy of Quantum Information Sciences, Beijing 100193, China}
% \affiliation{Beijing National Laboratory for Condensed Matter Physics, Institute of Physics, Chinese Academy of Sciences, Beijing 100190, China}
% \affiliation{University of Chinese Academy of Sciences, Beijing 100049, China}

\author{Meiling Li}
\thanks{These authors contributed equally to this work.}
\affiliation{Beijing Key Laboratory of Fault-Tolerant Quantum Computing, Beijing Academy of Quantum Information Sciences, Beijing 100193, China}

\author{Weijie Sun}
\affiliation{Beijing Key Laboratory of Fault-Tolerant Quantum Computing, Beijing Academy of Quantum Information Sciences, Beijing 100193, China}

\author{Yanjun Ma}
\affiliation{Beijing Key Laboratory of Fault-Tolerant Quantum Computing, Beijing Academy of Quantum Information Sciences, Beijing 100193, China}

\author{Zhenyu Mi}
\affiliation{Beijing Key Laboratory of Fault-Tolerant Quantum Computing, Beijing Academy of Quantum Information Sciences, Beijing 100193, China}

\author{Fei Yan}
\affiliation{Beijing Key Laboratory of Fault-Tolerant Quantum Computing, Beijing Academy of Quantum Information Sciences, Beijing 100193, China}

\author{Huikai Xu}
\affiliation{Beijing Key Laboratory of Fault-Tolerant Quantum Computing, Beijing Academy of Quantum Information Sciences, Beijing 100193, China}

\author{Pei Liu}
\email{liupei@baqis.ac.cn}
\affiliation{Beijing Key Laboratory of Fault-Tolerant Quantum Computing, Beijing Academy of Quantum Information Sciences, Beijing 100193, China}

\author{Guangming Xue}
\email{xuegm@baqis.ac.cn}
\affiliation{Beijing Key Laboratory of Fault-Tolerant Quantum Computing, Beijing Academy of Quantum Information Sciences, Beijing 100193, China}
\affiliation{Hefei National Laboratory, Hefei 230088, China}

\author{Haifeng Yu}
\email{hfyu@baqis.ac.cn}
\affiliation{Beijing Key Laboratory of Fault-Tolerant Quantum Computing, Beijing Academy of Quantum Information Sciences, Beijing 100193, China}
\affiliation{Hefei National Laboratory, Hefei 230088, China}

\begin{abstract}

Efficient qubit reset has become a key requirement for large-scale superconducting quantum processors, as initialization fidelity increasingly lags behind gate performance and quantum error correction demands rapid, high-fidelity state re-initialization.
This work proposes a compact on-chip dissipator consisting of two transmon dissipation modes and a thin-film resistor, and experimentally demonstrates an unconditional reset protocol.
With no additional control lines and negligible impact on the qubit $T_1$, the effective temperature of the transmon qubit is reduced from $49$~mK to $35$~mK.
Quantum process tomography is employed to benchmark the reset process, achieving a process error of $0.0028$.
This reset protocol is robust to pulse parameter deviations, thereby relaxing calibration requirements.
The dissipator architecture provides a practical, calibration-friendly reset solution for large-scale superconducting quantum processors.

\end{abstract}

\maketitle

\section{Introduction}

Superconducting circuits are one of the most promising platforms for universal quantum computation \cite{ladd2010quantum, preskill2018quantum}.
Despite decoherence times that still lag behind those of other platforms, superconducting systems satisfy several other requirements according to the DiVincenzo criteria \cite{divincenzo2000the}.
Recently, the single-qubit and two-qubit gate fidelities of superconducting transmon processors exceed $99.99\%$ \cite{li2023error, bland2025millisecond} and $99.9\%$ \cite{li2024realization, marxer2026above, zhang2026high}, respectively.
As gate errors decrease, errors in initialization, which is one of the DiVincenzo criteria, are emerging as a previously underappreciated performance bottleneck.
The simplest approach is natural cooling to the ground state.
However, this often requires waiting several times the relaxation time $T_1$, especially as the transmon $T_1$ increases; one either has to tolerate a longer cycle time or a larger initialization error.
Initialization errors already impair conventional readout error mitigation methods and consequently degrade the performance of several quantum applications \cite{guo2026false}.
Active reset protocols have therefore attracted considerable attention.
Moreover, quantum error correction demands rapid, high-fidelity qubit reset within each cycle \cite{chen2021exponential, he2025experimental}, making reset capability an increasingly important metric for superconducting processors.

Several active reset protocols have been demonstrated in superconducting systems over the past decade.
Quantum feedback \cite{riste2012initialization, vijay2012stabilizing, campagne2013persistent, salathe2018low} is one approach, in which the qubit state is first measured and a $\pi$-pulse is applied only when the excited state is detected.
This scheme is straightforward but suffers from measurement-induced state transitions \cite{sank2016measurement}, feedback latency, and the complexity of integrating low-latency control electronics at scale.
Another widely studied approach uses a dissipator, in which a dissipation channel carries the qubit excitation into the environment \cite{reed2010fast, murch2012cavity, geerlings2013demonstrating, magnard2018fast, egger2018pulsed, mcewen2021removing, zhou2021rapid, sunada2022fast, kim2025fast, huber2025parametric, chen2026fast}.
The temperature of the thermal bath surrounding the dissipator is of primary importance, as it determines whether the thermal excitation of the qubit can ultimately be suppressed.
Therefore, for a given thermal environment, the higher-frequency readout resonator tends to serve as the cold dissipator \cite{magnard2018fast, egger2018pulsed, zhou2021rapid, sunada2022fast, chen2026fast, huber2025parametric, omahen2026high}, although it suffers from a relatively complex and slow excitation transfer process.
More recently, evidence suggests that control lines in superconducting processors act as a considerable thermalization channel \cite{vaaranta2022dynamics, simbierowicz2024inherent}.
This finding offers new insight into the development of on-chip dissipators for qubit reset: on-chip components that reduce radiative thermal sources and enhance thermal conduction can serve as reset dissipators, with the ultimate goal of achieving high reset fidelity, fast reset speed, and low reset calibration complexity simultaneously.

This work proposes a compact on-chip dissipator comprising two transmon modes \cite{koch2007charge} and a thin-film resistor.
It features fast excitation swapping and dissipation, has negligible impact on the qubit $T_1$, and requires no additional control lines.
Based on this dissipator, an unconditional reset protocol with repeated reset cycles reduces the effective temperature of the qubit from $49$~mK to $35$~mK, and realizes a process fidelity of $0.9972$.
Even with deviations in the calibration parameters, the error rate of the reset process remains below $0.01$ for sufficiently large cycle numbers $n$.
These features make the reset protocol a practical solution for large-scale superconducting quantum processors.

\section{Device Design and Characterization}

\begin{figure}
    \centering
    \includegraphics{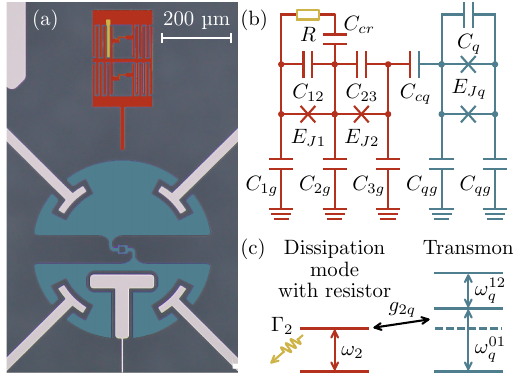}
    \caption{\textbf{Device design}.
    (a) False-colored micrograph and (b) the equivalent circuit of the superconducting transmon with its dissipator.
    The transmon qubit (blue), and the dissipator (red) with the resistor (yellow) are highlighted.
    (c) The energy-level diagram of the coupled transmon qubit and dissipation mode 2, with the coupling strength $g_{2q}$ and dissipation rate $\Gamma_2$.}
    \label{fig:1}
\end{figure}

The dissipator, which comprises a thin-film resistor and two transmon modes, is the core component of the reset architecture proposed in this work.
Figure~\ref{fig:1}(a) presents a false-colored micrograph of the dissipator unit together with the coupled transmon qubit, and Fig.~\ref{fig:1}(b) displays its equivalent circuit.
One of the dissipation modes acquires an excitation from the transmon through capacitive coupling, and subsequently releases this excitation via the resistor, thereby de-exciting the transmon.
The dissipation modes can in principle be realized using either a superconducting resonator or a transmon.
In the present design, two fixed-frequency transmons serve as the dissipation modes because they occupy a smaller footprint than resonators, and their compact structure is better suited to large-scale superconducting quantum processors.
The absorption rate of mode $i$ ($i=1,2$) is given by \cite{devoret1995quantum, cattaneo2021engineering}
\begin{equation}\label{eq:Gamma}
    \Gamma_i=\frac{R\lambda_i^{2}C_{ir}^{2}\omega_i\omega_c^{2}}{\hbar C_i^{2}(\omega_i^{2}+\omega_c^{2})}\left[\coth\left(\frac{\hbar\omega_i}{2k_{\rm B}T_e}\right)+1\right],
\end{equation}
where $\omega_i$ is the mode frequency, $\lambda_i\approx\sqrt{\hbar\omega_iC_i/2}$, $C_i$ is the total capacitance of mode $i$, $C_{ir}=C_{cr}$ is the coupling capacitance between mode $i$ and the resistor, $\omega_c$ is the effective RC cut-off frequency, $k_{\rm B}$ is the Boltzmann constant, and $T_e$ is the environmental bath temperature.
A detailed derivation of Eq.~\eqref{eq:Gamma} and the design parameters is provided in Section~I of the Supplementary Material.

Neglecting the dissipation induced by the resistor and retaining only the dominant capacitive interactions, the effective Hamiltonian of the coupled qubit-dissipator system reduces to
\begin{equation}\label{eq:H}
    \begin{aligned}
    \mathcal H&=\sum_{i=q,1,2}\left(\omega_i a_i^\dagger a_i+\frac{\alpha_i}{2}a_i^\dagger a_i^\dagger a_i a_i\right)\\
    &+\sum_{i=1,2}g_{iq}\left(a_i+a_i^\dagger\right)\left(a_q+a_q^\dagger\right),
    \end{aligned}
\end{equation}
where $\omega_i$, $\alpha_i$, and $a_i$ denote the frequency, anharmonicity, and annihilation operator of mode $i$ ($i=q,1,2$), respectively, and $g_{iq}$ is the coupling strength between dissipation mode $i$ and the qubit.

During the reset operation, the qubit frequency $\omega_q$ is tuned into resonance with one of the dissipation modes.
To avoid frequency crowding with neighboring components, the higher-frequency mode $\omega_2$ is selected as the primary dissipation channel.
The relevant energy levels and transitions are illustrated in Fig.~\ref{fig:1}(c).
The simplified model includes the transmon states $|0\rangle$, $|1\rangle$, and $|2\rangle$, together with the ground state $|g\rangle$ and first-excited state $|e\rangle$ of the dissipation mode $2$.

\begin{figure*}
    \centering
    \includegraphics{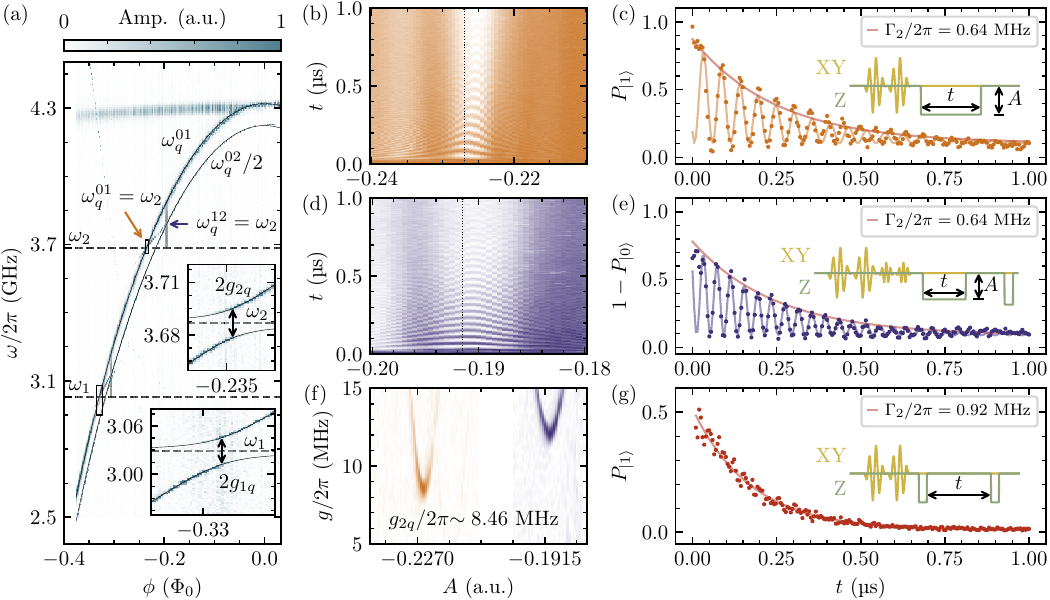}
    \caption{\textbf{System spectroscopy}.
    (a) Transmon spectrum showing the $|0\rangle\leftrightarrow|1\rangle$ transition frequency $\omega_q^{01}$ and the two-photon $|0\rangle\leftrightarrow|2\rangle$ transition $\omega_q^{02}/2$ as functions of the flux $\phi$.
    Two distinct dissipation modes $\omega_1$ and $\omega_2$ are indicated by black dashed lines.
    The insets magnify the anti-crossing regions marked by the black squares, showing splittings $2g_{2q}$ at $\omega_2$ and $2g_{1q}$ at $\omega_1$.
    Two coupling configurations are examined.
    First, $\omega_q^{01}$ crosses $\omega_2$ (i.e., $|1g\rangle\leftrightarrow|0e\rangle$), marked by the orange arrow.
    (b) iSWAP oscillations versus flux bias $A$ and duration $t$ for this configuration.
    (c) Damped Rabi oscillation at the dotted bias, yielding a decay rate $\Gamma_2/2\pi=0.64$~MHz; inset, the pulse sequence.
    Second, $\omega_q^{12}$ crosses $\omega_2$ (i.e., $|2g\rangle\leftrightarrow|1e\rangle$), marked by the purple arrow.
    The corresponding iSWAP dynamics and damped oscillation are shown in (d) and (e); inset of (e), the pulse sequence, in which the transmon is first prepared in $|2\rangle$ and a calibrated iSWAP pulse is applied before readout.
    (f) Extracted coupling strengths $g$ for the two configurations, yielding $g_{2q}/2\pi\approx8.46$~MHz for the first configuration.
    The extracted values are consistent with the splitting in (a) and satisfy the expected $\sqrt{2}$ scaling relation.
    (g) Static dissipation rate of mode 2, $\Gamma_2/2\pi=0.92$~MHz, measured by swapping the excitation into mode 2 and back with two calibrated iSWAP pulses separated by a variable delay.}
    \label{fig:2}
\end{figure*}

Figure~\ref{fig:2}(a) presents the measured transmon spectrum, with $\omega_q^{01}$ and $\omega_q^{02}/2$ plotted as functions of the flux $\phi$.
Two anti-crossing regions are clearly visible in the $\omega_q^{01}$ curve, corresponding to resonant coupling with the two dissipation modes $\omega_1$ and $\omega_2$.
The two-photon transition $\omega_q^{02}/2$ likewise exhibits two anti-crossing regions, occurring where $\omega_q^{12}=\omega_i$, i.e., at the $|2g\rangle\leftrightarrow|1e\rangle$ resonance.
Figure~\ref{fig:2}(b) displays the iSWAP dynamics between $|1g\rangle$ and $|0e\rangle$.
As the qubit frequency $\omega_q^{01}$ is tuned across $\omega_2$, the excitation undergoes coherent Rabi oscillations between these two states.
Owing to the intrinsic dissipation of mode $2$, the oscillation is damped, as shown in Fig.~\ref{fig:2}(c).
Similarly, when $\omega_q^{12}$ crosses $\omega_2$, the $|2g\rangle\leftrightarrow|1e\rangle$ exchange occurs, and the corresponding dynamics are presented in Figs.~\ref{fig:2}(d) and \ref{fig:2}(e).
Unlike the $|1g\rangle\leftrightarrow|0e\rangle$ case, the pulse sequence for the $|2g\rangle\leftrightarrow|1e\rangle$ configuration requires preparing the transmon in $|2\rangle$ before the iSWAP interaction.
Additionally, a calibrated iSWAP pulse is applied prior to readout to map the qubit population back into the $\{|0\rangle,|1\rangle\}$ subspace.
The coupling strengths extracted from the damped iSWAP oscillations are summarized in Fig.~\ref{fig:2}(f), which is consistent with the splitting in the inset of Fig.~\ref{fig:2}(a).
The Rabi frequency of the $|2g\rangle\leftrightarrow|1e\rangle$ iSWAP is approximately $\sqrt{2}$ times that of $|1g\rangle\leftrightarrow|0e\rangle$, consistent with Eq.~\eqref{eq:H}.
The decay rates extracted from the two damped oscillations are comparable and both lower than the static dissipation rate measured in Fig.~\ref{fig:2}(g).
This suggests that using passive decay of mode 2 during an idle interval, rather than the continuous iSWAP, enables a faster reset process.
The measured device parameters are summarized in Section~I of the Supplementary Material.

A potential concern when coupling a dissipator to a transmon for the purpose of reset is the degradation of qubit coherence.
The theoretical Purcell limit \cite{purcell1946resonance, reed2010fast} on $T_1$ is derived in Section~I of the Supplementary Material.
With properly designed parameters, the transmon $T_1$ can remain essentially unaffected by the presence of the dissipation mode.
In this work, $T_1$ is characterized for six transmons coupled with individual dissipators at their sweet point, with detailed data provided in Section~II of the Supplementary Material.
Compared with the $T_1$ of two bare transmons on the same chip, no significant degradation is observed.
Meanwhile, for the transmon selected in this work, the $T_1$ at its sweet point is $(68\pm2)$~\textmu s (corresponding to a detuning of $\sim 634$~MHz from $\omega_2$), and decreases to $(49\pm3)$~\textmu s when the qubit is detuned by $\sim 177$~MHz.
$T_1$ is reduced by approximately $28\%$.
These results indicate that, with appropriate frequency arrangement, the dissipator has negligible impact on processor performance.

\section{Robust Reset Protocol and Benchmarking}

\begin{figure}
    \centering
    \includegraphics{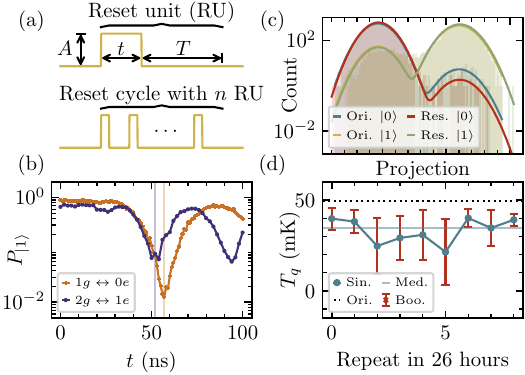}
    \caption{\textbf{Reset protocol}.
    (a) Schematic of the reset protocol.
    A reset unit (RU) comprises an iSWAP stage, which transfers the transmon excitation into the dissipator, followed by a delay stage in which the excitation decays.
    A reset cycle (RC) consists of $n$ repeated RU rounds.
    (b) Measured $P_{|1\rangle}$ as a function of the exchange duration $t$ for the two iSWAP configurations, used to calibrate the exchange duration.
    The vertical lines mark the selected durations, located at the minima corresponding to complete population transfer.
    (c) Single-shot readout histograms for the $|0\rangle$ and $|1\rangle$ states, compared without (original, Ori.) and with (reset, Res.) the RC applied before state preparation and measurement ($n=1$, $T=1$~\textmu s).
    The reduced separation between the original and reset $|0\rangle$ distributions reflects the suppression of initialization error.
    (d) Repeats of qubit effective temperature with neglected measurement error.
    The blue curve shows the single-shot result with the RC applied over 1024 shots per point (Sin.).
    The red markers show the same quantity estimated by bootstrapping, and the light blue line marks their median.
    The dotted line indicates the single-shot result without the RC (Ori.).
    }
    \label{fig:3}
\end{figure}

Building upon the dissipator, the reset protocol is illustrated in Fig.~\ref{fig:3}(a).
A reset unit (RU) consists of an iSWAP stage, followed by a dissipation stage.
In the iSWAP stage, the target iSWAP transition is selected by the flux bias $A$, requiring no measurement or feedback.
To maximize the effective dissipation rate, the qubit is returned to its idle frequency after the exchange rather than being held at the iSWAP resonance point.
The iSWAP duration and the dissipation waiting time are denoted by $t$ and $T$, respectively.
The duration $t$ is chosen as the time required for complete population transfer into the dissipation mode.
As shown in Fig.~\ref{fig:3}(b), the two iSWAP configurations require slightly different exchange durations for full population transfer.
Due to flux pulse distortion, the experimentally selected duration is slightly longer than the value predicted from the Rabi frequency shown in Fig.~\ref{fig:2}(f).

The full reset protocol employs a reset cycle (RC) composed of $n$ repeated RUs, referred to as $n$-RC, as depicted in Fig.~\ref{fig:3}(a).
Figure~\ref{fig:3}(c) demonstrates the reset performance in terms of the state preparation and measurement error of the transmon state $|0\rangle$.
Owing to the finite-temperature environment, a transmon exhibits residual thermal excitation.
When the transmon is prepared in $|0\rangle$ and subsequently measured, the probability $P(1|0)$ of obtaining the outcome $|1\rangle$ contains contributions from both the initialization error and the measurement error.
The effect of the reset can thus be assessed by comparing $P(1|0)$ with and without the RC applied.
The readout histograms of $|0\rangle$ in Fig.~\ref{fig:3}(c) display a clear reduction in $P(1|0)$ from $1.60\%$ without the reset to $0.66\%$ with the reset, indicating that the RC successfully suppresses the thermal excitation of the transmon qubit.
Assuming a constant measurement error, the thermal excitation can be described by an effective temperature that includes a fixed offset due to the measurement error,
\begin{equation}
T_q = -\frac{\hbar \omega_q^{01}}{2 k_{\mathrm{B}}\cdot\operatorname{arctanh}(2P(1|0)-1)}.
\end{equation}
Figure~\ref{fig:3}(d) presents $T_q$ over several repetitions after calibration of the reset process.
After accounting for sampling errors, the result reconstructed via the bootstrapping method also outperforms that obtained by passive relaxation alone.
The $T_q$ of the transmon qubit is lower under reset conditions, with a median of approximately $35$~mK, compared to $49$~mK without reset.
The reset employed for Fig.~\ref{fig:3}(c) is the $|1g\rangle\leftrightarrow|0e\rangle$ iSWAP with $A=-0.227$, $n=1$, $t=57$~ns, and $T=1$~\textmu s.
The RC employed for Fig.~\ref{fig:3}(d) spans various parameter combinations, all of which exhibit a pronounced reset effect.

It is particularly noteworthy that the success of the reset scheme implies a lower thermal population of the dissipator, especially in the case where the dissipator frequency is lower than the qubit frequency.
This indicates that the effective temperature of the dissipator without any control line is significantly lower than that of the qubit.
Assuming that, at the end of multiple dissipation cycles, the qubit and the dissipator reach the same thermal excitation (the same $P(1|0)$ value), the dissipator temperature $T_d$ can be estimated at approximately $30$~mK ($T_d/T_q=\omega_2/\omega_q$).
In large-scale superconducting quantum processors, thermal excitation introduced by control lines deserves particular attention.

Quantum process tomography (QPT) is employed to provide a more rigorous characterization of the reset operation.
For a non-unitary target process such as reset, a generalized expression,
\begin{equation}
    \mathcal F=\frac{\mathrm{Tr}\left(\chi\chi_{\rm ideal}\right)}{\mathrm{Tr}\left(\chi_{\rm ideal}\,\chi_{\rm ideal}\right)}
\end{equation}
is adopted, which maps the fidelity range from $[0,\mathrm{Tr}(\chi_{\rm ideal}\chi_{\rm ideal})]$ to $[0,1]$ while preserving the result for unitary operations.
A detailed derivation is provided in Section~III of the Supplementary Material.
Figure~\ref{fig:4}(a) presents the QPT result for an RC with $A=-0.227$, $t=57$~ns, $T=250$~ns, and $n=10$, yielding a process error $\epsilon=1-\mathcal{F}=0.0028$.

Figures~\ref{fig:4}(b) and \ref{fig:4}(c) show the dependence of $\epsilon$ on the RC parameters.
In Fig.~\ref{fig:4}(b), $A=-0.227$ and $t=57$~ns are fixed, and $\epsilon$ is plotted versus $n$ for several values of $T$.
Under these conditions, the extent of excitation transfer is held fixed while the degree of dissipation varies with $T$.
During the first few rounds, residual excitation remaining in the dissipation mode coherently returns to the transmon, causing pronounced oscillations in the reset error.
As $n$ increases, the excitation is progressively dissipated and the reset error stabilizes to a low value.
A shorter $T$ leads to larger residual excitation and stronger oscillations.
The inset shows $\epsilon$ as a function of $T$ at a fixed $n=10$, indicating that more complete dissipation within each RU reduces the reset error; however, an excessively long $T$ exacerbates the cumulative thermal excitation over the entire RC, potentially degrading the reset performance.
Figure~\ref{fig:4}(c) compares $\epsilon$ for different values of $A$ and $t$ at a fixed $T=100$~ns, with the reference configuration corresponding to a nearly perfect iSWAP exchange.
Compared with the reference (nearly perfect exchange), imperfect exchange leaves residual excitation on the transmon qubit, yet the reset error remains low and stable at large $n$, although the error oscillations are more pronounced at small $n$.
Taken together, Figs.~\ref{fig:4}(b) and \ref{fig:4}(c) demonstrate that by increasing the number of RC rounds $n$, a low reset error can be reliably achieved despite imperfections in $A$, $t$, or $T$.
The $n$-RC protocol therefore exhibits strong parametric robustness, effectively relaxing the calibration requirements.

\begin{figure}
    \centering
    \includegraphics{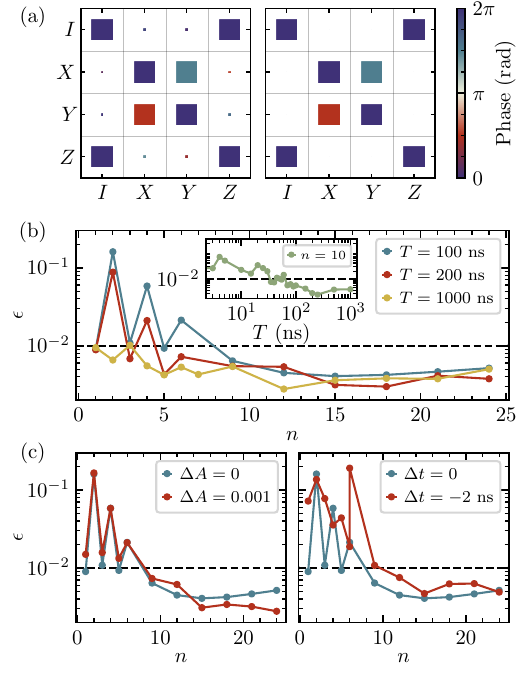}
    \caption{\textbf{Fidelity benchmarking}.
    (a) Process matrices $\chi$ of the experimental RC (left) and the ideal reset process (right), obtained via quantum process tomography (QPT) with $A=-0.227$, $t=57$~ns, $T=250$~ns, and $n=10$.
    The extracted process fidelity is $\mathcal{F}=0.9972$ (error $\epsilon=0.0028$).
    (b) QPT error $\epsilon$ as a function of the number of RC rounds $n$, plotted for several decay durations $T$.
    Inset, $\epsilon$ versus $T$ at a fixed round number $n=10$.
    Longer $T$ suppresses the residual excitation via more complete dissipation, whereas excessive duration exacerbates the accumulation of thermal excitation.
    (c) QPT error $\epsilon$ versus $n$ for imperfect iSWAP exchanges, with deviations in $A$ and $t$ at a fixed $T=100$~ns; the reference corresponds to a nearly perfect exchange.
    The multi-round RC protocol exhibits strong robustness against calibration imperfections.
    }
    \label{fig:4}
\end{figure}

\section{Summary and Outlook}

This work proposes a compact on-chip dissipator comprising two transmon modes and a thin-film resistor.
The dissipator design is systematically characterized through spectroscopic measurements, revealing two dissipation modes with coupling strengths consistent with the design values.
Building upon this architecture, an unconditional and robust reset protocol is implemented.
A total of no more than $2$~\textmu s reset process reduces the effective temperature of the transmon qubit from $49$~mK to $35$~mK. 
To further enhance the reset performance and robustness, an $n$-round reset cycle is introduced.
Quantum process tomography of a $10$-RC yields a process error of $\epsilon=0.0028$.
Systematic benchmarking further demonstrates that the $n$-RC protocol maintains, for sufficiently large $n$, a low reset error below $1\%$ under deliberate variations of the iSWAP amplitude $A$, the exchange duration $t$, and the dissipation waiting time $T$, confirming its strong parametric robustness.

The implementation of this scheme also highlights the importance of thermal management in a large-scale quantum processor.
First, it underscores the need to investigate the sources of thermal excitation of qubits in the $\sim10$~mK environment.
Second, this scheme also confirms that, for components that must remain cold, such as the dissipator, proper thermal management can indeed lower the effective temperature without a higher intrinsic frequency.

This design is extensible to multi-qubit architectures, where each qubit may be equipped with a dedicated dissipator or share one among several neighboring qubits through time-multiplexed control.
Several technical challenges remain open for future investigation, such as reducing the reset duration through optimized pulse shaping, analyzing the dissipator's thermal environment, and mitigating its impact on transmon decoherence.
The thin-film-resistor dissipator provides a new practical building block for next-generation large-scale superconducting quantum processors, combining fast operation, high fidelity, and intrinsic robustness against parameter deviations.

\section{Acknowledgments}

We thank Dr.~Weiyang Liu, Dr.~Yulong Li, and Dr.~Yirong Jin for helpful discussions.
We acknowledge the support from the Smart Grid-National Science and Technology Major Project (No.~2024ZD0803301),
the National Natural Science Foundation of China (No.~12604750, No.~92365206, No.~92565301),
and the Innovation Program for Quantum Science and Technology (No.~2021ZD0301802).

\bibliographystyle{apsrev4-2}
\bibliography{ref}

\clearpage
\onecolumngrid
\setcounter{figure}{0}
\setcounter{table}{0}
\setcounter{equation}{0}
\renewcommand{\thefigure}{S\arabic{figure}}
\renewcommand{\thetable}{S\arabic{table}}
\renewcommand{\theequation}{S\arabic{equation}}

\section{Circuit Quantization}

The main text presents the scanning electron micrograph and the equivalent circuit diagram of the device.
The full circuit consists of a transmon qubit and a dissipator, where the dissipator contains two modes, one of which is directly coupled to the thin-film resistor through a capacitor.
To describe the dissipation induced by the resistor in the quantum circuit, the resistor is replaced by an LC oscillator \cite{devoret1995quantum} and its coupling strength to the other resonant modes is computed.
The Lagrangian of the full system is written as
\begin{equation}
    \mathcal{L}=\mathcal{T}-\mathcal{U}=\frac{1}{2}\dot{\bm{\phi}}^{\,\mathrm{T}}[C]\,\dot{\bm{\phi}}-\mathcal{U},
\end{equation}
where $\mathcal{T}$ and $\mathcal{U}$ denote the kinetic and potential energies, respectively, and
\begin{equation}
    \mathcal{U}=E_{J1}[1-\cos(\phi_2-\phi_1)]+E_{J2}[1-\cos(\phi_3-\phi_2)]+\frac{L_r}{2}(\phi_4-\phi_1)^{2}+E_{Jq}[1-\cos(\phi_5-\phi_6)].
\end{equation}
The circuit contains six active nodes, and the node flux vector in the flux basis is defined as $\bm{\phi}^{\,\mathrm{T}}\equiv[\phi_1,\phi_2,\phi_3,\phi_4,\phi_5,\phi_6]$.
The capacitance matrix takes the form
\begin{fontsize}{10pt}{6pt}
\begin{equation}
    [C]=\begin{pmatrix}
        C_{12}+C_{1g}+C_{r} & -C_{12} & 0 & -C_{r} & 0 & 0 \\
        -C_{12} & C_{12}+C_{23}+C_{2g}+C_{cr} & -C_{23} & -C_{cr} & 0 & 0 \\
        0 & -C_{23} & C_{23}+C_{3g}+C_{cq} & 0 & -C_{cq} & 0 \\
        -C_{r} & -C_{cr} & 0 & C_r+C_{cr} & 0 & 0 \\
        0 & 0 & -C_{cq} & 0 & C_q+C_{qg}+C_{cq} & -C_q \\
        0 & 0 & 0 & 0 & -C_q & C_q+C_{qg}
    \end{pmatrix}.
\end{equation}
\end{fontsize}
To recover the relevant transmon degrees of freedom, a coordinate transformation is applied with the transformation matrix
\begin{equation}
    [S]=\begin{pmatrix}
        1 & 1 & 1 & 1 & 0 & 0 \\
        -1 & 1 & 0 & 0 & 0 & 0 \\
        0 & -1 & 1 & 0 & 0 & 0 \\
        -1 & 0 & 0 & 1 & 0 & 0 \\
        0 & 0 & 0 & 0 & 1 & -1 \\
        0 & 0 & 0 & 0 & 1 & 1
    \end{pmatrix}.
\end{equation}
The transformed coordinate vector reads
\begin{equation}
    \bm{\psi}=[S]\,\bm{\phi}=
    \begin{pmatrix}
        \phi_1+\phi_2+\phi_3+\phi_4 \\
        \phi_2-\phi_1 \\
        \phi_3-\phi_2 \\
        \phi_4-\phi_1 \\
        \phi_5-\phi_6 \\
        \phi_5+\phi_6
    \end{pmatrix}=
    \begin{pmatrix}
        \psi_0 \\ \psi_1 \\ \psi_2 \\ \psi_r \\ \psi_q \\ \psi_{qc}
    \end{pmatrix}.
\end{equation}
The transformed capacitance matrix is $\widetilde{[C]}=[S]^{-1}[C][S]$.
Since $\psi_0$ and $\psi_{qc}$ lack corresponding potential terms in $\mathcal{U}$, they constitute decoupled center-of-mass degrees of freedom that can be discarded, reducing the system to an effective $4\times4$ Lagrangian in the remaining variables.

The generalized momenta are obtained from the Lagrangian as $Q_i=\partial\mathcal{L}/\partial\dot\psi_i$, which can be written in matrix form as $\bm{Q}=(\Phi_0/2\pi)\,\widetilde{[C]}\,\dot{\bm{\psi}}$.
Applying the Legendre transformation $\mathcal{H}=\sum_i\dot\psi_i Q_i-\mathcal{L}$ yields the classical Hamiltonian
\begin{equation}
    \mathcal{H}=\frac{1}{2}\bm{Q}^{\,\mathrm{T}}\widetilde{[C]}^{-1}\bm{Q}+\mathcal{U},
\end{equation}
where
\begin{equation}
    \widetilde{[C]}^{-1}=
    \begin{pmatrix}
        \frac{1}{\widetilde{C}_1} & \frac{\widetilde{C}_{12}}{\widetilde{C}_1\widetilde{C}_2} & \frac{\widetilde{C}_{1r}}{\widetilde{C}_1\widetilde{C}_r} & \frac{\widetilde{C}_{1q}}{\widetilde{C}_1\widetilde{C}_q} \\
        \frac{\widetilde{C}_{12}}{\widetilde{C}_1\widetilde{C}_2} & \frac{1}{\widetilde{C}_2} & \frac{\widetilde{C}_{2r}}{\widetilde{C}_2\widetilde{C}_r} & \frac{\widetilde{C}_{2q}}{\widetilde{C}_2\widetilde{C}_q} \\
        \frac{\widetilde{C}_{1r}}{\widetilde{C}_1\widetilde{C}_r} & \frac{\widetilde{C}_{2r}}{\widetilde{C}_2\widetilde{C}_r} & \frac{1}{\widetilde{C}_r} & 0 \\
        \frac{\widetilde{C}_{1q}}{\widetilde{C}_1\widetilde{C}_q} & \frac{\widetilde{C}_{2r}}{\widetilde{C}_2\widetilde{C}_q} & 0 & \frac{1}{\widetilde{C}_q}
    \end{pmatrix}.
\end{equation}

In the present design, the coupling capacitances $C_{cr}$ and $C_{cq}$ are much smaller than all other capacitances, and the ground capacitances satisfy $C_{1g},C_{2g},C_{3g}\ll C_{12},C_{23}$.
Under these experimentally realistic conditions, the inverse capacitance matrix elements simplify to
\begin{align}
    \widetilde{C}_1 &\approx C_{12}+\frac{C_{1g}(C_{2g}+C_{3g})}{C_{1g}+C_{2g}+C_{3g}}, &
    \widetilde{C}_2 &\approx C_{23}+\frac{C_{3g}(C_{1g}+C_{2g})}{C_{1g}+C_{2g}+C_{3g}}, \\
    \widetilde{C}_r &\approx C_{r}, &
    \widetilde{C}_q &\approx C_{q}+\frac{C_{qr}}{2}, \\
    \widetilde{C}_{12} &\approx -\frac{C_{1g}C_{3g}}{C_{1g}+C_{2g}+C_{3g}}, &
    \widetilde{C}_{1r} &\approx C_{cr}, \\
    \widetilde{C}_{2r} &\approx C_{cr}\frac{C_{1g}C_{3g}}{C_{12}(C_{1g}+C_{2g}+C_{3g})}, &
    \widetilde{C}_{1q} &\approx C_{cq}\frac{C_{1g}}{2(C_{1g}+C_{2g}+C_{3g})}, \\
    \widetilde{C}_{2q} &\approx C_{cq}\frac{C_{1g}+C_{2g}}{2(C_{1g}+C_{2g}+C_{3g})}.
\end{align}
It should be noted that the resistor-coupled mode is treated here solely for the purpose of extracting the coupling strength between the dissipator and the resistor, and is not itself treated as an independent resonant mode.

Quantizing the circuit and adopting the Duffing approximation for the transmon, the Hamiltonian becomes
\begin{align}
    \mathcal{H}&=\sum_{i}\frac{Q_{i}^{2}}{2C_{i}}-E_{Ji}\cos(\psi_{i})+\sum_{i\neq j}\frac{C_{ij}}{C_{i}C_{j}}Q_{i}Q_{j}\notag\\
    &=\sum_{i}\omega_i a_i^{\dagger}a_i+\frac{1}{2}\alpha_i a_i^{\dagger}a_i^{\dagger}a_i a_i+\sum_{i\neq j}g_{ij}(a_i+a_i^{\dagger})(a_j+a_j^{\dagger}),
\end{align}
where the indices $i,j$ run over $\{1,2,q\}$.
The mode frequencies, anharmonicities, and coupling strengths are given by $\omega_i=\sqrt{8E_{Ci}E_{Ji}}-E_{Ci}$, $\alpha_i=-E_{Ci}$, $g_{ij}=C_{ij}\sqrt{\omega_i\omega_j}/(2\sqrt{C_iC_j})$, with the charging energy $E_{Ci}=e^{2}/(2C_i)$.

The dissipation rate can be precisely tuned through the choice of the resistance $R$ and the coupling capacitance.
The mode decay induced by capacitive coupling to a resistor has been rigorously formulated, yielding the emission and absorption rates \cite{cattaneo2021engineering}
\begin{equation}
    \Gamma^{\downarrow}=\frac{R\lambda_A^{2}C_{c}^{2}\omega_A\omega_c^{2}}{\hbar C_{\Sigma}^{2}(\omega_A^{2}+\omega_c^{2})}\left[\coth\!\left(\frac{\hbar\omega_A}{2k_{\rm B}T}\right)+1\right],
\end{equation}
\begin{equation}
    \Gamma^{\uparrow}=\frac{R\lambda_A^{2}C_{c}^{2}\omega_A\omega_c^{2}}{\hbar C_{\Sigma}^{2}(\omega_A^{2}+\omega_c^{2})}\left[\coth\!\left(\frac{\hbar\omega_A}{2k_{\rm B}T}\right)-1\right],
\end{equation}
where $\Gamma^{\downarrow}$ and $\Gamma^{\uparrow}$ are the emission and absorption rates, respectively, determined by the bath temperature $T$.
Here $\hbar$ is the reduced Planck constant, $k_{\rm B}$ is the Boltzmann constant, $R$ is the resistance, $C_{c}$ is the coupling capacitance, $C_{\Sigma}$ is the total qubit capacitance, and $\omega_A$ is the qubit frequency.
$\omega_c$ is the effective cut-off frequency, whose precise value depends on the coupling between the system and the environment.
The typical qubit operating frequency ($\sim5$\,GHz) lies far below the system cut-off, and $\omega_c/2\pi=1.0$\,THz \cite{koral2025decoherence} is adopted in the numerical calculations presented in this work.
For a transmon qubit, $\omega_A\approx(\sqrt{8E_JE_C}-E_C)/\hbar$ and $\lambda_A\approx\sqrt{\hbar\omega_AC_{\Sigma}/2}$.
The decay rate of dissipator mode $i$ ($i=1,2$) induced by the resistive coupling is
\begin{equation}
    \Gamma_i^{\downarrow}=\frac{R\lambda_i^{2}C_{ir}^{2}\omega_i\omega_c^{2}}{\hbar C_{i}^{2}(\omega_i^{2}+\omega_c^{2})}\left[\coth\!\left(\frac{\hbar\omega_i}{2k_{\rm B}T}\right)+1\right].
\end{equation}
At the base temperature of the dilution refrigerator ($T\approx 10$\,mK), the spontaneous absorption rate $\Gamma_i^{\uparrow}$ is negligible.

Table~\ref{tab:design_parameters} lists the design parameters of the device.
Table~\ref{tab:simulation_parameters} compares the simulated frequencies and coupling strengths with the experimentally measured values.
Good agreement is found between the designed and measured parameters.

\begin{table}
    \centering
    \caption{Design Parameters}\label{tab:design_parameters}
    \begin{tabular*}{\linewidth}{@{\extracolsep{\fill}}*{7}{w{c}{\dimexpr\linewidth/7-2\tabcolsep\relax}}}
        \toprule
            $C_{1g}$ (fF) & $C_{2g}$ (fF) & $C_{3g}$ (fF) & $C_{12}$ (fF) & $C_{23}$ (fF) & $C_{q}$ (fF) & $C_{qg}$ (fF) \\
        \midrule
            37 & 32 & 38 & 53 & 62 & 19 & 144 \\
        \toprule
            $C_{cq}$ (fF) & $C_{cr}$ (fF) & $E_{J1}/2\pi$ (GHz) & $E_{J2}/2\pi$ (GHz) & $E_{Jq}/2\pi$ (GHz) & $R$ ($\Omega$) & \\
        \midrule
            2.8 & 10 & 7.1 & 7.8 & 11.7 & 30 & \\
        \bottomrule
    \end{tabular*}
\end{table}

\begin{table}
    \centering
    \caption{Simulated and Measured Parameters}\label{tab:simulation_parameters}
    \begin{tabular*}{\linewidth}{@{\extracolsep{\fill}}*{8}{w{c}{\dimexpr\linewidth/8-2\tabcolsep\relax}}}
        \toprule
            &$\omega_q/2\pi$ (GHz) & $\omega_1/2\pi$ (GHz) & $g_{1q}/2\pi$ (MHz) & $\Gamma_1/2\pi$ (MHz) & $\omega_2/2\pi$ (GHz) & $g_{2q}/2\pi$ (MHz) & $\Gamma_2/2\pi$ (MHz)\\
        \midrule
            Simulated & 4.250 & 3.125 & 17.76 & 1.71 & 3.676 & 9.44 & 0.91 \\
            Measured & 4.309 & 3.029 & 16.60 & - & 3.687 & 8.46 & 0.92 \\
        \bottomrule
    \end{tabular*}
\end{table}

\section{Qubit Decoherence}

\begin{figure}
    \centering
    \includegraphics[width=0.5\linewidth]{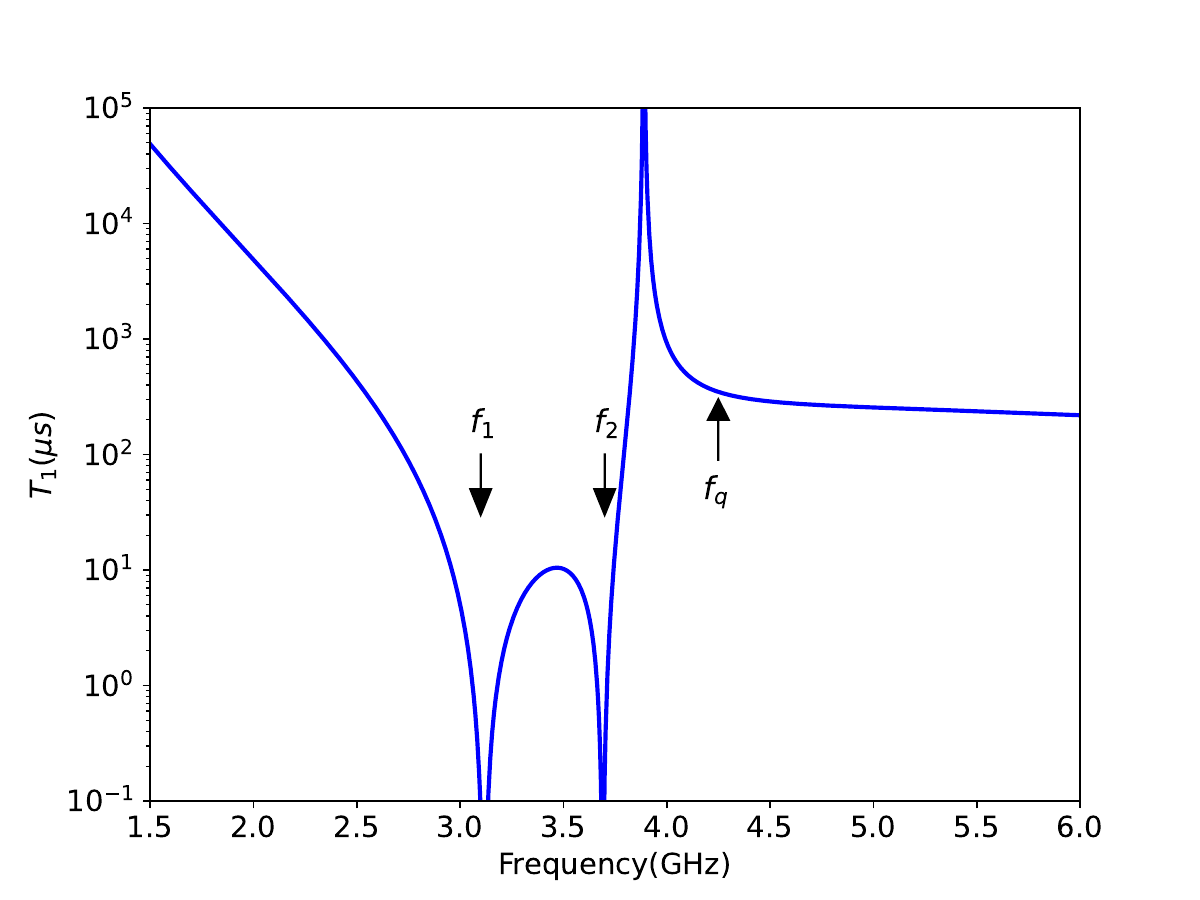}
    \caption{\textbf{Purcell limit of the transmon}.}
    \label{fig:purcell}
\end{figure}

\begin{figure*}
    \centering
    \includegraphics[]{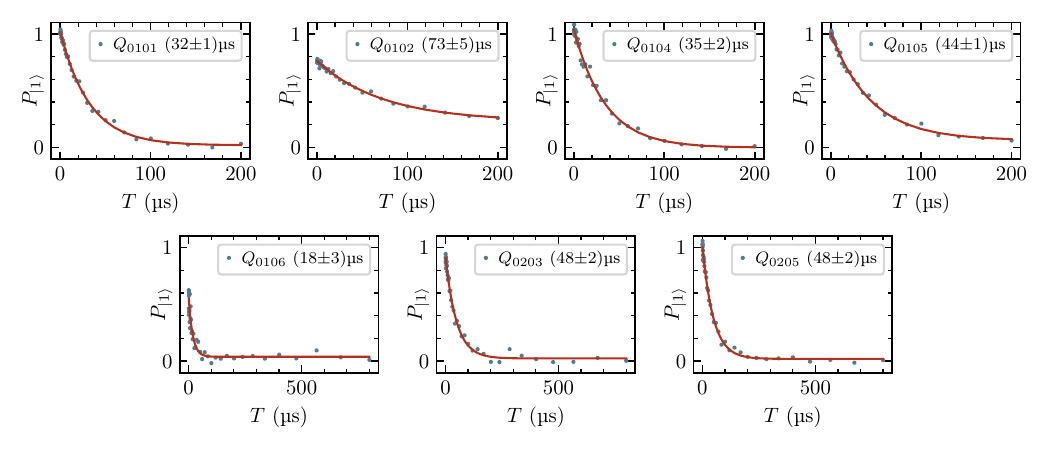}
    \caption{\textbf{Energy relaxation time for test transmon qubits}.}
    \label{fig:s2}
\end{figure*}

Employing a resistive element as the dissipative source allows the dissipation rate of the dissipator to be controlled without external microwave circuitry.
However, similar to a lossy resonator, the resistor also introduces a Purcell decay channel for the qubit \cite{purcell1946resonance}.
The Purcell-limited coherence time is given by \cite{reed2010fast}
\begin{equation}
    T_1^{\mathrm{Purcell}}=\frac{C_q}{\operatorname{Re}[Y(\omega_q)]},
\end{equation}
where $C_q$ is the qubit capacitance and $\operatorname{Re}[Y(\omega_q)]$ is the real part of the admittance presented by the environment at the qubit frequency.
Finite-element electromagnetic simulations are performed to evaluate this admittance, with the dissipator Josephson junctions replaced by lumped inductors and the resistor modeled as a thin-film resistive layer.
The simulation results are shown in Fig.~\ref{fig:purcell}.
At the designed qubit frequency of $4.25$\,GHz, the Purcell-limited $T_1$ exceeds $300$\,$\mu$s.
Furthermore, transmons without the dissipator structure were also fabricated on the same chip, and their measured $T_1$ values show no statistically significant difference from those of the dissipator-coupled transmons (see Fig.~\ref{fig:s2}).

The experiments reported in this work are actually carried out on a $6\times 12$ transmon qubit array.
The first six qubits on the first array are each equipped with an individual dissipator.
Table~\ref{tab:qubit_decorence} lists the $T_1$ values of these six qubits measured near their sweet points, together with those of two reference qubits fabricated without the dissipator structure.
No statistically significant difference is observed between the two groups.

\begin{table}
    \centering
    \caption{Qubit $T_1$}\label{tab:qubit_decorence}
    \begin{tabular*}{\linewidth}{@{\extracolsep{\fill}}*{8}{w{c}{\dimexpr\linewidth/8-2\tabcolsep\relax}}}
        \toprule
            $Q_{0101}$ & $Q_{0102}$ & $Q_{0103}$ & $Q_{0104}$ & $Q_{0105}$ & $Q_{0106}$ & $Q_{0203}$ & $Q_{0205}$\\
        \midrule
            ($32\pm1$) \textmu s & ($73\pm5$) \textmu s & ($68\pm2$) \textmu s & ($35\pm2$) \textmu s & ($45\pm1$) \textmu s & ($18\pm3$) \textmu s & ($48\pm2$) \textmu s & ($48\pm2$) \textmu s\\
        \bottomrule
    \end{tabular*}
\end{table}

\begin{figure*}
    \centering
    \includegraphics[]{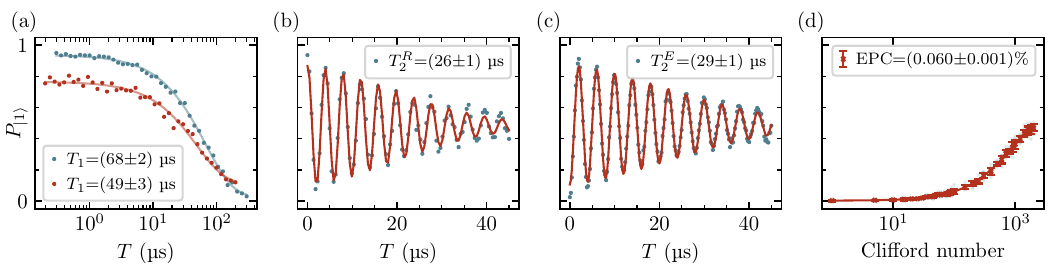}
    \caption{\textbf{Decoherence and single-qubit gate fidelity}.
    (a) The measured $T_1$ at the sweet point (blue) and at $A=-0.2$ (red).
    (b) The Ramsey interference and (c) spin-echo results.
    (d) The randomized benchmarking result for the single-qubit gate with the error per Clifford (EPC) of $(0.060\pm0.001)\%$.}
    \label{fig:s1}
\end{figure*}

The experiments were performed on qubit $Q_{0103}$.
Figure~\ref{fig:s1} presents the corresponding decoherence characterization.
Notably, the $T_1$ of this qubit at its sweet point is $(68\pm2)$~\textmu s (corresponding to a detuning of $\sim 634$\,MHz from $\omega_2$), and decreases to $(49\pm3)$~\textmu s when the qubit is detuned by $\sim 177$\,MHz from the dissipation mode $\omega_2$.
The single-qubit gate fidelity is characterized by randomized benchmarking \cite{knill2008randmoized}, yielding a single-Clifford error rate of $(0.060\pm0.001)\%$, much lower than the initialization and measurement error rates.

\section{Quantum Process Tomography for the Non-Unitary Process}

In quantum process tomography (QPT), a quantum process can be expressed as
\begin{equation}
    \mathcal{E}(\rho)=\sum_{ij}\chi_{ij}P_{i}\rho P^{\dagger}_{j},
\end{equation}
where the super-operator $\chi$ satisfies the following conditions,
\begin{equation}
    \begin{cases}
        \chi=\chi^{\dagger},\\
        \operatorname{Tr}(\chi)=1,\\
        \forall v, |\chi-vI|=0,\ 0\leq v\leq 1.
    \end{cases}
\end{equation}
Given the normalized non-identity generators $\{\lambda_{k}|k=1,2,\cdots,d^{2}-1\}$ of $\operatorname{SU}(d)$, the matrix $\chi$ can be expanded as
\begin{equation}
    \chi=\frac{1}{d}I+\bm{r}\cdot\bm{\lambda},
\end{equation}
where $\bm{\lambda}=(\lambda_{1},\lambda_{2},\cdots,\lambda_{d^{2}-1})^{\mathrm {T}}$, $\bm{r}=r\bm{n}=(r_{1},r_{2},\cdots,r_{d^{2}-1})^{\mathrm{T}}\in\mathbb{R}^{d^{2}-1}$, and $|\bm{n}|=1$. 
The trace condition $\operatorname{Tr}(\chi)=1$ is automatically satisfied. 
Furthermore, the purity-like quantity $\operatorname{Tr}(\chi^{2})$ bounds the length of the generalized Bloch vector,
\begin{equation}
    \operatorname{Tr}\left(\chi^{2}\right)=\frac{1}{d}+r^{2}\leq 1\Rightarrow r\leq \sqrt{\frac{d-1}{d}},
\end{equation}
which restricts $\chi$ to a compact region within the space of admissible process matrices.
The Hilbert-Schmidt inner product between two process matrices $\chi$ and $\chi^{\prime}$ provides a natural distance measure
\begin{equation}
    \operatorname{Tr}\left(\chi\chi^{\prime}\right)=\frac{1}{d}+\bm{r}\cdot\bm{r}^{\prime}.
\end{equation}
By the Cauchy-Schwarz inequality,
\begin{equation}
    -|\bm{r}|\cdot|\bm{r}^{\prime}|\leq\bm{r}\cdot\bm{r}^{\prime}\leq|\bm{r}|\cdot|\bm{r}^{\prime}|.
\end{equation}
When $\bm{r}^{\prime}=k\bm{r}$ for some $k\in\mathbb{R}$, the Cauchy-Schwarz equality is attained and the two process matrices are said to be collinear in the generalized Bloch representation.
The case $k=0$ corresponds to $\chi^{\prime}=I/d$ (the maximally mixed process), for which $\operatorname{Tr}(\chi\chi^{\prime})=1/d$ and the fidelity bounds hold trivially.
Therefore, $k\neq 0$ is considered in what follows.
Under this condition, the eigenvalues $\{v^{\prime}\}$ of $\chi^{\prime}$ are related to the eigenvalues $\{v\}$ of $\chi$ through the characteristic equation
\begin{equation}
    \begin{aligned}
        |\chi^{\prime}-v^{\prime}I| &=\left|\frac{1}{d}I+k\bm{r}\cdot\bm{\lambda}-v^{\prime}I\right| \\
        &=k^{d}\left|\frac{1}{d}I+\bm{r}\cdot\bm{\lambda}-\left(\frac{v^{\prime}}{k}+\frac{1}{d}-\frac{1}{dk}\right)I\right| \\
        &=k^{d}\,|\chi-vI|=0.
    \end{aligned}
\end{equation}
Thus,
\begin{equation}
    v=\frac{v^{\prime}}{k}+\frac{1}{d}-\frac{1}{dk}\Leftrightarrow v^{\prime}=kv+\frac{1-k}{d}=\left(v-\frac{1}{d}\right)k+\frac{1}{d}.
\end{equation}
Given that $\sum v=1$, the trace of $\chi^{\prime}$ evaluates directly as
\begin{equation}
    \sum v^{\prime}=k\sum v+\frac{1-k}{d}\cdot d=k\cdot 1+1-k=1,
\end{equation}
confirming that the transformed process matrix remains trace-normalized.

The expression for $v^{\prime}$ in terms of $v$ and $k$ allows one to translate the eigenvalue positivity constraints from $\chi$ to $\chi^{\prime}$.
Since $k$ controls the relative scaling between the two process matrices, bounding $k$ directly yields bounds on the Hilbert-Schmidt inner product $\operatorname{Tr}(\chi\chi^{\prime})$.
These bounds are derived below separately for positive and negative $k$.

\subsection{Case $k>0$: Upper Bound}

When $\chi$ possesses a zero eigenvalue ($\exists v=0$),
\begin{equation}
    0\leq v^{\prime}=\frac{1-k}{d}\leq 1\Rightarrow 1-d\leq k\leq 1.
\end{equation}
Since $\bm{r}\cdot\bm{r}^{\prime}=k|\bm{r}|^{2}$, the upper bound $k\leq 1$ yields $\bm{r}\cdot\bm{r}^{\prime}\leq|\bm{r}|^{2}$, and therefore
\begin{equation}
    \operatorname{Tr}\left(\chi\chi^{\prime}\right)=\frac{1}{d}+\bm{r}\cdot\bm{r}^{\prime}\leq\frac{1}{d}+|\bm{r}|^{2}=\operatorname{Tr}\left(\chi^{2}\right).
\end{equation}
If instead all eigenvalues of $\chi$ are strictly positive ($\forall v>0$), then there exists at least one eigenvalue $v\leq 1/d$ (since $\sum v=1$). From the condition $v^{\prime}\geq 0$,
\begin{equation}
    v^{\prime}=\left(v-\frac{1}{d}\right)k+\frac{1}{d}\geq 0\Rightarrow k\leq\frac{1}{1-dv}.
\end{equation}
Since $v\leq 1/d$ implies $1/(1-dv)\geq 1$, the constraint $k\leq 1$ obtained from the zero-eigenvalue case is always the tighter bound, and the strictly positive sub-case therefore does not affect the upper bound on $\operatorname{Tr}(\chi\chi^{\prime})$.

\subsection{Case $k<0$: Lower Bound}

For any admissible eigenvalue set $\{v\}$ except the maximally mixed case (where $\bm{r}=\bm{0}$ and all eigenvalues equal $1/d$), there exists a maximum eigenvalue $v_{\max}>1/d$. The positivity condition $v^{\prime}\geq 0$ yields
\begin{equation}
    v^{\prime}=\left(v-\frac{1}{d}\right)k+\frac{1}{d}\geq 0\Rightarrow k\geq-\frac{1}{dv_{\max}-1}.
\end{equation}
Substituting the bound on $k$ and using $\bm{r}\cdot\bm{r}^{\prime}=k|\bm{r}|^{2}$,
\begin{equation}
    \operatorname{Tr}\left(\chi\chi^{\prime}\right)=\frac{1}{d}+k|\bm{r}|^{2}\geq\frac{1}{d}-\frac{|\bm{r}|^{2}}{dv_{\max}-1}.
\end{equation}
Expressing $|\bm{r}|^{2}=\operatorname{Tr}(\chi^{2})-1/d=\sum v^{2}-1/d$,
\begin{equation}
    \begin{aligned}
        \operatorname{Tr}\left(\chi\chi^{\prime}\right) &\geq\frac{1}{d}-\frac{\sum v^{2}-1/d}{dv_{\max}-1}
        =\frac{v_{\max}}{dv_{\max}-1}-\frac{\sum v^{2}}{dv_{\max}-1}.
    \end{aligned}
\end{equation}
Since $\sum v=1$ and $0\leq v_i\leq v_{\max}$, the convex combination satisfies $\sum v^{2}\leq v_{\max}\sum v=v_{\max}$, and therefore $v_{\max}-\sum v^{2}\geq 0$.
With $dv_{\max}>1$ the denominator is positive, so $\operatorname{Tr}(\chi\chi^{\prime})\geq 0$.

\subsection{Summary}

Combining the constraints from the $k>0$ and $k<0$ cases, the Hilbert-Schmidt inner product is bounded by
\begin{equation}
    0\leq\operatorname{Tr}\left(\chi\chi^{\prime}\right)\leq\operatorname{Tr}\left(\chi^{2}\right),
\end{equation}
provided that the ideal process matrix $\chi$ possesses at least one zero eigenvalue.
This condition is satisfied by any process that maps the qubit Hilbert space to a lower-dimensional subspace, such as a reset operation that projects arbitrary input states onto the ground state.
Dividing through by $\operatorname{Tr}(\chi^{2})$ then yields the normalized fidelity measure
\begin{equation}
    \mathcal{F}(\chi,\chi^{\prime})=\frac{\operatorname{Tr}(\chi\chi^{\prime})}{\operatorname{Tr}(\chi^{2})} \in [0,1],
\end{equation}
which coincides with the standard process fidelity for unitary operations while remaining well-defined for non-unitary processes.
This expression is used in the main text to quantify the reset fidelity from the experimentally reconstructed process matrix.

\subsection{Initialization and Measurement Errors in QPT}

In the single-qubit setting considered in this work, initialization errors do not propagate through entangling operations, so the conventional readout-correction method remains valid.
Moreover, the same reset process is applied prior to both the readout correction calibration and the QPT measurement sequence, ensuring that the initialization and readout errors are consistent throughout the experiment.
Under these conditions, both sources of error are small, and the readout correction does not introduce a systematic bias into the reconstructed Pauli expectation values.
The readout-correction method is therefore applicable to the present experimental configuration.

\end{document}